\documentclass[fleqn,twoside]{article}
\usepackage{espcrc2}

\usepackage{graphicx}
\usepackage[figuresright]{rotating}

\newcommand{\AmS}{{\protect\the\textfont2
  A\kern-.1667em\lower.5ex\hbox{M}\kern-.125emS}}

\title{Neutrino Properties and Tests of Symmetries}

\author{Andr\'e de Gouv\^ea\address{Department of Physics and Astronomy, Northwestern University, \\ 2145 Sheridan Road, Evanston IL, 60208-3112, USA}}
       
\begin{document}

\begin{abstract}
The fact that neutrinos have mass allows them to possess many other properties, like a magnetic dipole moment and a finite decay width. Theoretical expectations for these are, however, well beyond current experimental bounds, and the discovery of new exotic neutrino properties would indicate the presence of (more) new physics. Here, I review some of the current bounds on exotic neutrino properties.

Neutrino experiments are unique probes of several fundamental space-time symmetries, including invariance under CP and Lorentz transformations. I discuss how precision neutrino experiments  are used to test some of the most fundamental principles of physics at unprecedented levels.
\vspace{1pc}
\end{abstract}

\maketitle

\section{Introduction}

Our picture of the high energy world changed with the amazing discovery that neutrinos have tiny, yet nonzero, masses. This na\"{\i}vely simple fact requires a qualitative modification of the Standard Model. 

We currently do not know what the ``New Standard Model'' Lagrangian is. There are two distinct ``minimal'' candidates: 
\begin{equation}
{\cal L}={\cal L}_{\rm old~SM}-\lambda^{\alpha\beta}\frac{L_{\alpha}HL_{\beta}H}{2M}+h.c.  \label{Lmaj} 
\end{equation}
or
\begin{equation}
{\cal L}={\cal L}_{\rm old~SM}+i\bar{N_i}\slash\!\!\!\partial N_i-\lambda^{\alpha i}\bar{L}_{\alpha}HN_i + h.c.. \label{Ldirac}
\end{equation}
Here, $L_{\alpha}$ are the left-handed lepton weak isodoublets, $H$ is the Higgs scalar weak isodoublet, $N_i$ are right-handed fermion gauge singlets, $\lambda$ are dimensionless coefficients, and $M$ is a (very large) mass parameter. ${\cal L}_{\rm old~SM}$ is the Lagrangian of the ``old'' Standard Model, which predicts that neutrinos are strictly massless. 

In the case of Eq.~(\ref{Lmaj}), our description of Nature is no longer valid up to arbitrarily high energy scales (or even the Planck scale), but breaks down at some energy scale close to $M$. If $M\gg\langle H\rangle$, the smallness of the neutrino mass is promptly understood: $m_{\nu}=\lambda\langle H\rangle^2/M\ll \langle H\rangle$. $\langle H\rangle={\cal O}(10^2~\rm GeV)$ is the characteristic scale of all fundamental charged fermion masses (up to a dimensionless factor, which varies between $10^{-5}$ and $1$). Finally, the Lagrangian Eq.~(\ref{Lmaj}) predicts that the neutrinos are Majorana fermions and that lepton number minus baryon number is not strictly conserved in Nature.  

In the case of Eq.~(\ref{Ldirac}), only renormalizable operators are added to the old Standard Model, and there is no concrete reason to believe that this description of Nature is not valid all the way up to energy scales where gravitational interactions become relevant. There is, however, no natural mechanism for understanding why neutrino masses are so small -- $m_{\nu}=\lambda\langle H\rangle$, and the current data  require $\lambda^{\alpha i}={\cal O}(10^{-12})$ ({\it cf.}~the electron Yukawa coupling, which is ${\cal O}(10^{-5})$). Finally, Eq.~(\ref{Ldirac}) predicts that the neutrinos are Dirac fermions, like all of its charged counterparts. It is important to note that Eq.~(\ref{Ldirac}) is {\sl not} the most general renormalizable Lagrangian consistent with all gauge symmetries of the the Standard Model -- the relevant operator $M^{ij}\bar{N}^c_iN_j$ is absent. This absence needs to be ``enforced'' by, for example,  imposing a new symmetry on the Standard Model, such as a global $U(1)_{B-L}$. 

Except for the faith of baryon number minus lepton number, which will hopefully be unambiguously revealed by searches for neutrinoless double beta decay \cite{beta_talk}, both candidate Lagrangians share a remarkable feature: except for endowing the neutrinos with nonzero masses, they do not lead to any other observable physics effects!

The statuses of current (still fruitless) searches for effects that violate the predictions of Eqs.~(\ref{Lmaj}) and (\ref{Ldirac}) are discussed in the next section. Neutrino tests of the validity of a few space-time symmetries, some expected to be violated, like CP-invariance, others fundamental to our description of Nature, like CPT-invariance, are discussed in section 3.  A brief summary and conclusions follow.

\section{Neutrino Exotic Properties}

The fact that neutrinos have mass, combined with the unprecedented abundance of precise neutrino data, allows one to probe whether the neutrinos are endowed with other unexpected properties. These include an anomalously large magnetic and/or electric dipole moment or an anomalously short lifetime. The confirmation of either would signify that the minimal guesses for Nature's Lagrangian, Eq.~(\ref{Lmaj}) and Eq.~(\ref{Ldirac}), are incomplete. 

\subsection{Neutrino Dipole Moments}

Like all other massive fermions, neutrinos are expected to have nonzero electric and magnetic dipole moments, $\mu_{\nu}$. The nature of $\mu_{\nu}$ will depend on whether the neutrinos are Majorana or Dirac fermions:
\begin{equation}
{\cal L}_{m.m.}^{\rm Maj}=\mu_{\nu}^{ij}(\nu_i\sigma_{\mu\nu}\nu_jF^{\mu\nu})+h.c., 
\end{equation}
where $\mu_{\nu}^{ij}=-\mu_{\nu}^{ji}$, or
\begin{equation}
{\cal L}_{m.m.}^{\rm Dirac}=\mu_{\nu}^{ij}(\bar{\nu}_i\sigma_{\mu\nu}N_j F^{\mu\nu}) + h.c..  
\end{equation}
In either version of the New Standard Model, the elements of $\mu_{\nu}$ are predicted to be tiny \cite{mu_SM}: 
\begin{equation}
\mu_{\nu}^{ij}\leq
\frac{3eG_F}{8\sqrt{2}\pi^2}m_{\nu}=3\times 10^{-20}\mu_B\left(\frac{m_{\nu}}{10^{-1}~\rm eV}\right),
\end{equation}
where $\mu_B=e/2m_e$ is the Bohr magneton.

The current bounds on linear combinations of elements of $\mu_{\nu}$ are provided by a large variety of experimental probes, including 
\begin{itemize}
\item precision studies of $\bar{\nu}_{e}e^-\to\nu_{\beta}~(\bar{\nu}_{\beta})~e^-, ~\forall \beta$ ($\beta=e,\mu,\tau$) \cite{mu_talk};
\item searches for a solar electron antineutrino flux. A nonzero flux would arise due to neutrino interactions with the Sun's magnetic field, plus the effects of neutrino oscillations \cite{mu_solar};
\item various astrophysical probes, including the cooling rate of red-giant stars \cite{Raffelt}. 
\end{itemize}
Current 90\% confidence level bounds of $\mu_{\nu}<1.0\times 10^{-10}\mu_B$ are widely accepted \cite{PDG}, while studies
of neutrinos from astrophysical objects (including our Sun) hint at  $\mu_{\nu}<O(10^{-[12\div11]}\mu_B)$ \cite{mu_solar,Raffelt}.
 
Na\"{\i}vely, one would expect other new electroweak-scale physics to yield ``large'' contributions to the neutrino dipole moments,
\begin{equation}
\mu\sim \frac{e h^2}{M_{\rm new}^2}m_{f},
\end{equation}
where $m_f$ is a charged fermion mass, $h$ a dimensionless coupling and $M_{\rm new}$ the new physics scale. Hence, searches for neutrino magnetic moments constrain the new physics scale and coupling, similar to searches for rare charged lepton process, like $\mu\to e\gamma$ \cite{Savoy}, and precision measurements of the anomalous magnetic moment of the muon.

\subsection{Neutrino Decays}

Now that we know the neutrinos have mass, we expect the two heavier neutrinos to decay into lighter objects. Indeed, only the lightest neutrino is currently guaranteed to be stable, given that it is the lightest known fermion.

In the New Standard Model, There are two allowed neutrino decay modes: $\nu_i\to\nu_j\gamma$ or 
$\nu_i\to\nu_j\nu_k\nu_l$ (here $\nu$ stands for neutrinos and/or antineutrinos). The electromagnetic neutrino decay is governed by the same type of magnetic moment operator discussed in the previous subsection, and expectations for the neutrino lifetime are absurdly long: $\tau>10^{38}$~years for $m_{\nu}\sim 1$~eV \cite{decay_SM}. Similarly, $\tau(\nu\to 3\nu)>10^{39}$~years \cite{decay_SM}. Needless to say, current experimental upper bounds are many orders of magnitude away from the New Standard Model expectations. For example, 
constraints on $\nu\to\nu\gamma$ from failed searches for magnetic dipole moments impose the following bound on the neutrino ``electromagnetic'' lifetime \cite{bound_mu_lifetime}:
\begin{equation}
\tau>5\times 10^{11} \left(\frac{10^{-10}\mu_B}{\mu_{\nu}}\right)^2~{\rm years},
\end{equation}
for $m_{\nu}\sim1$~eV.

The observation of a finite neutrino lifetime, similar to that of a neutrino dipole moment, would indicate the presence of physics beyond the New Standard Model. These are either ``bread-and-butter'' $1/M_{\rm new}$ effects or, more interestingly, indicate the existence of new, very light yet  still unobservable, degrees of freedom (say, new quasi-massless scalars or pseudo-scalars).

Experimental bounds on the neutrino lifetime depend tremendously on the decay mode. Model independent bounds are extraordinarily weak, and depend on the neutrino mass hierarchy and the elements of the leptonic mixing matrix. Arguably, the best bounds are for the $\nu_2$-lifetime, and come from solar neutrino data \cite{BeacomBell} -- and are very easy to estimate. Given that one astronomical unit is roughly 500 light-seconds, the absence of neutrino-decay effects in the solar neutrino data bound, very roughly,
\begin{eqnarray} 
&\gamma\tau>500~{\rm s};  \\
&\tau>500\frac{m}{E}~{\rm s}\sim 10^{-4}~{\rm s}\left(\frac{m}{1~eV^2}\right)\left(\frac{5~\rm MeV}{E}\right). 
\end{eqnarray}
Significantly better (by many others of magnitude) bounds are expected from the future detection of astrophysical neutrinos, as discussed, for example, in recent studies of the detection of very high energy astrophysical neutrinos \cite{Beacometal} and of relic supernova neutrinos \cite{Lisisupernova}.

\section{Tests of Fundamental Symmetries}

Neutrino experiments are capable of exploring with unmatched precision whether several fundamental symmetries, including CP/T-invariance, Lorentz invariance, and CPT-invariance, are indeed respected in the leptonic sector. 
   
\subsection{CP-Invariance and T-Invariance}   

Given that there are at least three neutrino species, CP-invariance violation and T-invariance violation are expected. Furthermore, assuming that the same mechanism for CP-invariance violation that operates in the quark sector is at work in the leptonic sector, one expects that all lepton number conserving, CP-invariance violating phenomena should be parametrized by the same physical parameter, namely the ``Dirac'' CP-odd phase $\delta$ of the leptonic mixing matrix.\footnote{Throughout, I use the PDG parametrization of the leptonic mixing matrix \cite{PDG}.}

Currently, there is no positive signal for CP-invariance violation in the neutrino sector. Next generation neutrino experiments, however, are being planned to probe for CP-invariance violating effects and, most importantly, to determine the mechanism for CP-invariance violation in the leptonic sector. Very schematically, the several search channels include (i) comparing $P(\nu_{\alpha}\to\nu_{\beta})\times P(\bar{\nu}_{\alpha}\to\bar{\nu}_{\beta})$ for various $\alpha, \beta$, which probes the validity of CP-invariance (in practice, the usefulness of this comparison depends on how well the matter effects are known), and (ii) comparing $P(\nu_{\alpha}\to\nu_{\beta})\times P(\nu_{\beta}\to\nu_{\alpha})$ for various $\alpha, \beta$, which  probes the validity of T-invariance. For more details, see \cite{Petcov-talk}.
 
 If neutrinos are Majorana fermions, the CKM mechanism predicts other sources of CP-invariance violation. These are controlled by the so-called ``Majorana'' CP-odd phases, and are observable in lepton number violating processes. For example, very precise measurements of the lifetime for neutrinoless double beta decay, combined with futuristic computations of the relevant nuclear matrix elements and measurements of the neutrino oscillation parameters, may allow one to determine one linear combination of Majorana phases \cite{Maj_bb}. Curiously enough, while the CP-odd phase can be measured in this way, its effect is not CP-invariance violating \cite{Maj_CP}. There are several {\it bona fide} CP-invariance violating phenomena mediated by Majorana phases (see, for example, \cite{Maj_CP}). These, however, are exceedingly hard to observe, given that the rate for all (lepton number violating) observables involved are directly proportional to the very tiny neutrino masses.
 
 Finally, in the see-saw model \cite{seesaw}, there are several other CP-odd parameters that control potentially observable CP-invariance violating phenomena. These are related, for example, to CP-invariance violation in the decay of the super heavy ``right-handed'' neutrinos into leptons and Higgs bosons. While these phenomena cannot be observed today (or any time in the foreseeable future), they may have been ubiquitous in the very early, very hot stages of the Universe, and may have played a fundamental role in generating the currently observed baryon asymmetry of the Universe \cite{leptogenesis}.

\subsection{Lorentz Invariance}

Lorentz invariance is one of the pillars of the quantum field theories used to describe all microscopic phenomena observed to date. The experimental discovery that Lorentz invariance is violated would radically change our picture of the Universe we live in.

Violation of Lorentz invariance in the neutrino sector can manifest itself as a modified dispersion relation for the neutrino states: $E^2-|\vec{p}|^2\neq m^2$. Modified neutrino dispersion relations lead to deviations of the characteristic $L/E$ dependency of the neutrino oscillation probability, and can hence be looked for in precise long-baseline neutrino experiments.

Here, I'll describe one simple formalism, discussed in the literature by several authors \cite{Lorentz_viol,Lorentz_mine}, and will illustrate how neutrino oscillations bound Lorentz invariant violating effects. From within this formalism several interesting phenomenological possibilities arise, including the possibility of explaining all neutrino data via Lorentz invariance violation alone (without any neutrino masses) \cite{Kos_Mewes}! Here, I only discuss straight forward bounds on sub-leading effects due to the violation of Lorentz invariance.  

Consider adding to the Standard Model the following Lagrangian:
\begin{equation}  
{\cal L}_{\rm Lorentz}\supset A^{ij}_{\mu}\bar{\nu_i}\gamma^{\mu}\nu_j+B^{ij}_{\mu\nu}\bar{\nu}\sigma^{\mu\nu}\nu + H.c.,
\end{equation}
where $A^{ij}$ ($B^{ij}$) are a Lorentz vectors (tensors) and $i,j=1,2,3$. Lorentz invariance is ``spontaneously broken'' if $A$ and/or $B$ have nonzero vacuum expectation values. Let us assume that in a convenient reference frame $\langle A^{\mu}_{ij}\rangle=(V_{ij}/2,\vec{0})$. Under these circumstances, and in the limit $E,|\vec{p}|\gg m, V$,
\begin{equation}
E=\vec{p}+\frac{m^2}{2|\vec{p}|}\pm V.
\label{Lviol}
\end{equation}  
The plus (minus) sign applies for neutrinos (antineutrinos), meaning that in this formalism CPT-invariance is also violated. Note that, in Eq.~(\ref{Lviol}), $V$ plays the same role as the well-known matter potential, which modifies the propagation of neutrinos in matter. For this reason, $V$ can be referred to as the matter potential for neutrino propagation in a Lorentz invariant violating ``ether.''   

Two-flavor ether oscillations are given by 
\begin{eqnarray}
& P(\nu_e\to\nu_x)=\sin^22\theta_{\rm eff}\sin^2\left(\frac{\Delta_{\rm eff}}{2}L\right),  \\
& \Delta_{\rm eff}^2=\left(\Delta\cos2\theta-V\right)^2+\left(\Delta\sin2\theta+V_{ex}\right)^2,  \\
& \Delta_{\rm eff}\sin2\theta_{\rm eff}=\Delta \sin2\theta+V_{ex}, \\
& \Delta_{\rm eff}\cos2\theta_{\rm eff}=\Delta \cos2\theta-V_{ex}, 
\end{eqnarray}
where $\Delta=\Delta m^2/2E$, $V=2(V_{ee}-V_{xx})$ (I use the notation of \cite{Lorentz_mine}). For antineutrinos, the same expressions apply, with $V_{ij}\to-V_{ij}$. In the presence of the ether, neutrinos and antineutrinos have (different) energy dependent ``mixing angles'' and oscillation frequencies.

Current neutrino data set bounds for $V_{ij}$. Very conservatively, the absence of ether effects will bound $V_{ij}<\Delta m^2/2E$, in such a way that one can estimate $V_{ex}<10^{-6}$~eV$^2$/MeV  (from solar and KamLAND data) and $V_{\mu x}<10^{-3}$~eV$^2$/GeV (from the atmospheric data). Hence, current oscillation data bound all $V_{ij}<10^{-21}$~GeV (at least). The precise values of these upper bounds can be found in the literature  \cite{Lorentz_bounds}.

\subsection{CPT-invariance}

CPT-invariance is a consequence of the fact that our description of Nature is {\sl local, causal} and {\sl Lorentz invariant}. In the previous subsection, I discussed a model where CPT-invariance was violated through the violation of Lorentz invariance.  Here, I discuss another manifestation of CPT-invariance violation: the possibility that neutrino and antineutrino masses are different.

\begin{figure}[htb]
\vspace{9pt}
\includegraphics[width=18pc]{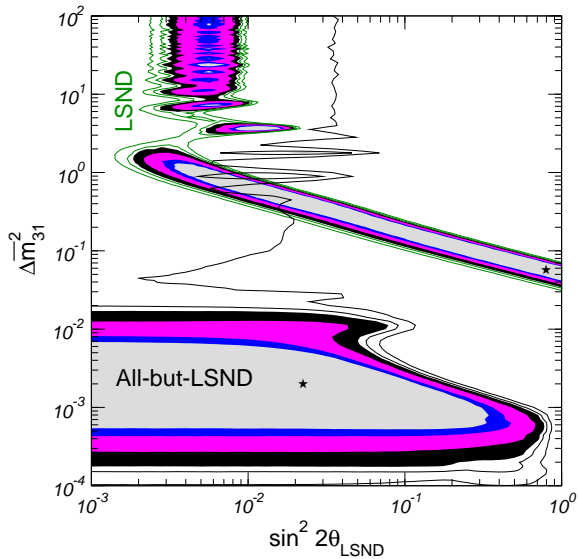}
\caption{90\%, 95\%, 99\%, and three sigma confidence level allowed regions (filled) in the $\Delta \bar{m}^2_{13} = \Delta m^2_{\rm LSND}\times\sin^22\theta_{\rm LSND}$ plane required to explain the LSND signal, together with the corresponding allowed regions from a global analysis \cite{GG_M_S} of all-but-LSND data. The contour lines correspond to   $\Delta \chi^2 =13$ and 16 (3.2 
and 3.6 sigma, respectively). See \cite{GG_M_S} for details.}
\label{LSND_CPT}
\end{figure}

The possibility that $m_{\nu}\neq m_{\bar{\nu}}$ was raised in \cite{Murayama_Yanagida} in order to address a slight inconsistency of the neutrino data from Supernova 1987A and to resolve the LSND anomaly \cite{LSND} without the addition of new degrees of freedom. Currently, these types of solution to the LSND anomaly \cite{LSND_CPT} are experimentally disfavored, because
(i) KamLAND (antineutrinos) and solar (neutrinos) data require small, consistent mass-squared differences -- $\Delta m^2_{\rm sol}\simeq\Delta\bar{m}^2_{\rm KamLAND}\ll \Delta \bar{m}^2_{\rm LSND}$;\footnote{Here, $\bar{m}$ and $\bar{\theta}$ refer to masses and mixing angles in the antineutrino sector, which may differ from $m$ and $\theta$, associated to the neutrino sector.}
(ii) The atmospheric neutrino data require that the mass-squared difference that governs both $\nu_{\mu}\leftrightarrow\nu_{\tau}$ {\sl and} $\bar{\nu}_{\mu}\leftrightarrow\bar{\nu}_{\tau}$ transitions 
be much smaller than the mass-squared difference required to solve the LSND anomaly -- $\Delta \bar{m}^2_{\rm atm}\ll \Delta \bar{m}^2_{\rm LSND}$ at more than the three sigma confidence level, as depicted in figure \ref{LSND_CPT}. See \cite{GG_M_S} for details. 

Since CPT-invariance violation is currently disfavored by the neutrino oscillation data, I proceed to discuss how the neutrino data can be used to constrain several observables that serve to parametrize CPT-invariance violation. In particular, I'll discuss bounds on $\Delta(\Delta m_{ij}^2)\equiv|\Delta m_{ij}^2-\Delta \bar{m}_{ij}^2|$ and $\Delta(\sin^2\theta_{ij})\equiv|\sin^2\theta_{ij}-\sin^2\bar{\theta}_{ij}|$.

\begin{figure*}[htb]
\vspace{9pt}
\includegraphics[width=38pc]{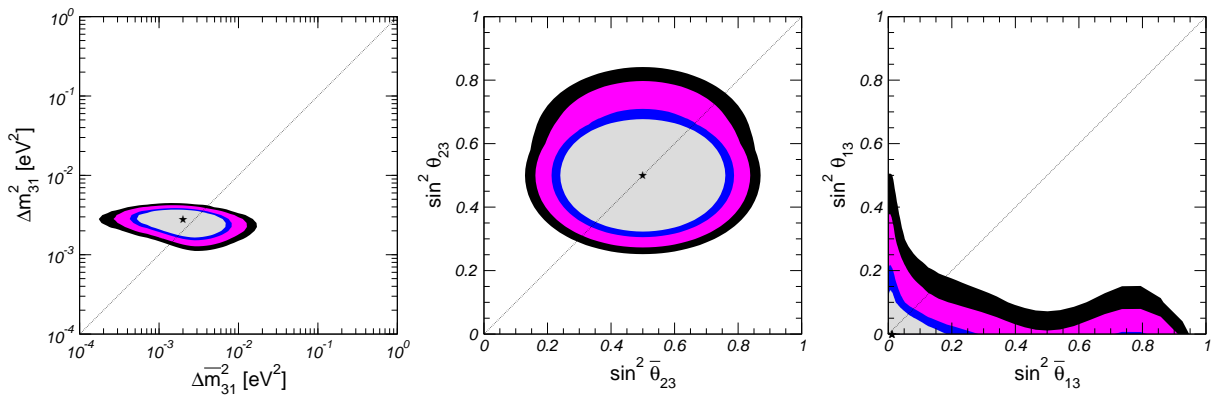}
\caption{Allowed regions for the largest neutrino and anti-neutrino mass splittings ($\Delta m^2_{13}$ and $\Delta \bar{m}^2_{13}$) and mixing angles  ($\theta_{23}$ and  $\bar{\theta}_{23}$, $\theta_{13}$ and  $\bar{\theta}_{13}$) \cite{GG_M_S}. The different contours correspond to the two-dimensional allowed regions at 90\%, 95\%, 99\%, and three sigma confidence levels from
all-but-LSND data. The best fit point is marked with a star. See \cite{GG_M_S} for details.}
\label{delta_delta_atm}
\end{figure*}

Figure \ref{delta_delta_atm} depicts the values of $\Delta m^2_{13}$, $\sin^2\theta_{23}$, $\sin^2\theta_{13}$, $\Delta \bar{m}^2_{13}$, $\sin^2\bar{\theta}_{23}$, and $\sin^2\bar{\theta}_{13}$, 
allowed by the current atmospheric neutrino data \cite{GG_M_S}. From the figure, it is easy to estimate
\begin{eqnarray}
&\Delta(\Delta m^2_{13})<1.1\times 10^{-2}~\rm eV^2,  \label{ddm_atm} \\
&\Delta (\sin^2\theta_{23})<0.45, \label{dds_atm}
\end{eqnarray}
at the three sigma confidence level (see also \cite{Kearns}). Only modest improvements are expected from the MINOS experiment \cite{MINOS}, given that it will measure precisely $\Delta m^2_{13}$ but not $\Delta \bar{m}^2_{13}$, whose uncertainty currently dominates the bound above. An antineutrino long-baseline experiment or a next generation atmospheric neutrino experiment (especially one capable of distinguishing neutrinos from antineutrinos) is required to significantly improve on Eqs.~(\ref{ddm_atm},\ref{dds_atm}).

\begin{figure*}[htb]
\vspace{9pt}
\includegraphics[width=38pc]{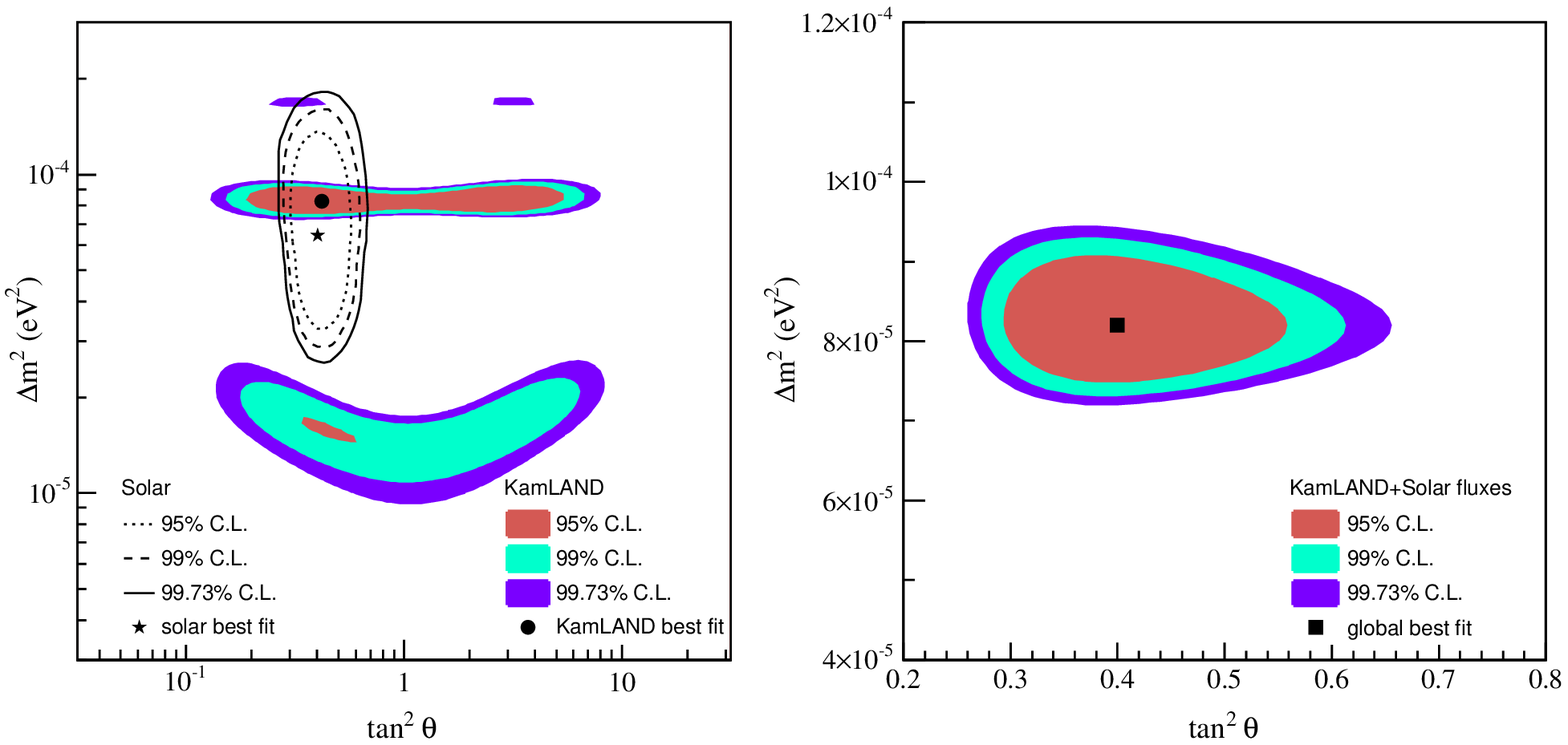}
\caption{(a) Allowed regions of neutrino oscillation parameters from KamLAND anti-neutrino data (shaded regions) and solar neutrino experiments (lines) \cite{KamLAND}. (b) Result of a combined two-neutrino oscillation analysis of KamLAND and the observed solar neutrino fluxes under
the assumption of CPT invariance. See \cite{KamLAND} for details.}
\label{delta_delta_sol}
\end{figure*}

More stringent limits can be obtained in the ``1--2'' sector by comparing KamLAND and solar data. Figure \ref{delta_delta_sol}(a) depicts the region of two-flavor $\nu_e\leftrightarrow\nu_x$ parameter space allowed by the study of the disappearance of electron antineutrinos at KamLAND and of electron neutrinos at several solar neutrino experiments \cite{KamLAND}. It is very impressive that several, very different neutrino probes point to the same region of parameter space. This agreement allows one to combine neutrino and antineutrino data with confidence and obtain the very precisely measured values of $\Delta m^2_{12}$ and $\sin^2\theta_{12}$ depicted in Fig.~\ref{delta_delta_sol}(b).   

Here, instead, we use the apparent KamLAND-solar agreement to bound CPT-invariance violation, and obtain \cite{AdG-CPG}: 
\begin{eqnarray}
&\Delta(\Delta m^2_{12})<1.1\times 10^{-4}~\rm eV^2, \label{ddm_sol} \\
&\Delta (\sin^2\theta_{12})<0.6, \label{dds_sol}
\end{eqnarray}
at the three sigma confidence level. The bound Eq.~(\ref{ddm_sol}) is currently dominated by the uncertainty on $\Delta m^2_{12}$, measured with the solar data. Modest improvements are expected from new data from SNO \cite{SNO}, while more significant improvements may come from a next-generation solar neutrino experiment. The bound Eq.~(\ref{dds_sol}) is dominated by the KamLAND data, which determines with only moderate precision $\sin^2\bar{\theta}_{12}$.  Worse than that, however, even if the KamLAND data were much more precise, one would not be able to do better than $\Delta (\sin^2\theta_{12})<|\cos2\bar{\theta}_{12}|$ \cite{AdG-CPG}. This is due to the fact that matter effects at KamLAND are very small, and one cannot tell whether $\bar{\theta}_{12}$ is on the dark side ($\bar{\theta}_{12}>\pi/4$). Improving on this situation may prove to be very difficult, even for next-next-generation long-baseline experiments. See \cite{AdG-CPG} for more details.

For completeness, I also mention that the current upper bound on $\Delta(\sin^2\theta_{13})<{\cal O}(1)$. We currently know, from the CHOOZ experiment \cite{CHOOZ}, that $\sin^22\theta_{13}<0.1$ (roughly), while $\sin^2\theta_{13}$ is loosely constrained by atmospheric and solar data (see, for example, figure \ref{delta_delta_atm}). Significant improvements are expected from next generation reactor and long-baseline experiments \cite{future_talks}. 

It is interesting to try to place the bounds obtained above in the larger arena of tests of Lorentz invariance violation \cite{PDG}. For example, assuming that $\Delta(\Delta m_{ij}^2)=|(m_j^2-\bar{m}_j^2)-(m_i^2-\bar{m}_i^2)|$ is representative of $|m_j^2-\bar{m}_j^2|$, which is quite reasonable (barring finely-tuned cancellations), Eq.~(\ref{ddm_sol}) represents the currently strongest bound on a particle--antiparticle mass-squred difference ({\it cf.} $|m^2_K-m^2_{\bar{K}}|=0.25$~eV$^2$ \cite{hitoshi_CPT}). 

\section{Summary, Conclusions} 

In the ``New Standard Model,'' neutrinos are not expected to possess observable exotic properties, like an electric dipole moment or a finite width. As far as it is concerned, it ``only'' remains to determine the values of all neutrino  masses and mixing parameters, whether CP-invariance (and T-invariance) is violated, and whether lepton number minus baryon number is conserved in all physical processes (already a very full research program!).

Nonetheless, massive neutrinos, combined with precision neutrino data, allow one to look for (more) new physics. Some neutrino processes serve as unique (albeit usually relatively weak) probes of $1/M_{\rm new}$ effects, not dissimilar from their charged lepton ``relatives'' (rare muon decays, electron and muon electric and magneitc dipole moments, etc). Other neutrino processes provide information on elusive light fermions and/or scalars, which may be inaccessible through other means.

Furthermore, neutrinos serve as narrow but very deep, unique probes of ``earth-shattering'' effects (Lorentz invariance violation, CPT-invariance violation, etc), that if observed would required a long and hard revision of some of the fundamental principles of physics. The unprecedented sensitivity comes from the ``quantum interferometric'' nature of the oscillation phenomena. It is always useful to keep in mind that neutrino oscillations have allowed us to observe the neutrino masses themselves, and that these may be a manifestation of physics at otherwise-inaccessible energy scales.

\section*{Acknowledgments}

It is a pleasure to thank the organisers of the Neutrino 2004 conference for the invitation to present this talk and  for putting together a well-organized and very stimulating meeting.

\end{document}